\documentstyle[12pt]{article}
\catcode`\@=11 \@addtoreset{equation}{section}  
   
\title{Superrelativity as a unification of quantum
theory and relativity (II)}
\author{P. Leifer}
\date{Mortimer and Raymond Sackler Institute of Advanced Studies \\ Tel-Aviv University, Tel-Aviv 69978, Israel \\
e-mail:leifer@ccsg.tau.ac.il}
\begin{document}
\maketitle
\begin{abstract}
A underlying dynamical structure for both relativity
and quantum theory---``superrelativity'' has been 
proposed on order to overcome the well known 
incompatibility between these theories.

The relationship between curvature of spacetime
(gravity) and curvature of the projective Hilbert
space of pure quantum states is established as well.

\end{abstract}
\section{Introduction. About ``Superrelativity''}
A new principle of ``superrelativity'' (SuperR) has 
already been
discussed in my previous reports [1,2]. In the framework
of this principle a non-linear 
equation of motion for relativistic scalar field
(see (6.23) in [1]) was established. In this work we 
will study the physical meaning of this equation on 
the basis of an approxipate solution.

A few words about general properties of our approach.
Notions of {\it material point, event, and classical
spacetime} in both special (SR) and general relativity
(GR) are liable to lead to confusion at the quantum
level. Insead of these obsolete objects we use a new 
set of primordial elements. Namely, they are {\it pure quantum state, quantum transition, and quantum state
space}, respectively. In the framework of our model the 
{\it fundamental scalar field} is rendered in a self-
interacting non-linear field configuration---``droplet''.
The proper surrounding field of the droplet is a 
non-abelian (relative to the transformation group of 
the Fourier components of the scalar field) gauge 
field of the connection in the complex projective 
Hilbert space of pure quantum states CP(N-1). The 
principle of ``superequivalence'' {\it identifies
this unified gauge field with the real physical fields
of non-local elementary particles}. That is 
the ``superequivalence'' principle establishes a 
parallelism
between GR and SuperR. This parallelism means that in
GR the {\it freely falling frame} serves for the 
description
of motion of the material point. In SuperR a {\it local
functional frame} connected with proper components of the 
scalar field configuration serves for the description 
of the evolution of the {\it quantum state in the
unified 
surrounding gauge field of the connection in CP(N-1)}.

The equivalence principle of Einstein is based on the
experimental fact that acceleration of bodies in the
gravitation field is independent on masses of these
bodies [3]. This situation is physically equivalent
to the motion of the system of bodies in accelerated 
frame. We can not, of course, put the {\it criterion of  
identical acceleration} in the basis of geometrization
of quantum physical fields.  In quantum field theory the
notion of ``acceleration'' is poor at best and ambiguous
at worst because quantum particle has some internal
structure. Furthermore, at a deeper level there is
{\it entanglement  and even indistinguishability of
``internal'' and ``external'' degrees of freedom}.
Therefore in quantum regime we can not act 
{\it literally} as Einstein in GR but only in his 
{\it spirit} [1,2].  

We have put at the basis of our 
``superequivalence'' principle the fact that in {\it all 
interactions of quantum (``elementary'') particles  
there is a conservation law of electric charge}.
Then group of isotropy of a pure quantum state $|\Psi>$
is $H=U(1)_{el} \times U(N-1)$. For ordinary Hilbert 
space C(N) the variations of a pure quantum state 
$|\Psi>$ lie in
the coset $G/H=SU(N)/S[H=U(1)_{el} \times U(N-1)]$. 
It is clear that variations of the pure 
quantum state are due to some physical interaction;
the effect of the interaction has the geometric 
structure of a coset, 
i.e. the structure of the complex projective Hilbert
space CP(N-1) [4]
\begin{equation} 
G/H=SU(N)/S[U(1)_{el} \times U(N-1)]=CP(N-1).
\end{equation}
This statement has a general character and does not 
depend on particular properties of the pure quantum
state. The reason  for the change of motion of material
point is the existence of a force. The reason for the 
change of a pure quantum 
state is interaction which may be modeled by unitary transformations from the coset (1.1). The reaction 
of a material point is acceleration. The reaction
of a pure quantum state is the deformation of the 
``ellipsoid of polarisation'' [1,2]. One-parameter
transformations from the coset create the geodesic
flow which are defined by the matrix $\hat{T}(\tau,g)$
(5.7) [1]. Therefore geodesics in CP(N-1) play an 
important but quite different role than geodesics in GR [1,2].

In the local coordinates
\begin{equation}
\pi^i_{(0)}=\Psi^i/\Psi^0      
\end{equation}
one can build a local frame for which
\begin{equation}
D_\sigma(\hat{P})=\Phi^i_\sigma (\pi,P)\frac{\delta}
{\delta \pi^i}
+\Phi^{i*}_\sigma (\pi,P)\frac{\delta}{\delta \pi^{i*}}
\end{equation}
spans the tangent Hilbert space relative
to the Fubini-Study metric 
\begin{equation}
G_{ik*}=2 \hbar R^2 {(R^2 + \sum_{s=1}^{N-1}|\pi^s|^2) \delta_{ik} 
-\pi^{i*} \pi^k \over (R^2+\sum_{s=1}^{N-1}|\pi^s|^2)^2}. 
\end{equation}
[1,2]. The coefficients these tangent vectors are 
defined by 
\begin{equation}
\Phi_{\sigma}^i(\pi;P_{\sigma}) = \lim_{\epsilon \to 0} \epsilon^{-1}
\biggl\{\frac{[\exp(i\epsilon P_{\sigma})]_m^i \Psi^m}{[\exp(i \epsilon P_{\sigma}]_m^0
\Psi^m }-\frac{\Psi^i}{\Psi^0} \biggr\}=
 \lim_{\epsilon \to 0} \epsilon^{-1} \{ \pi^i(\epsilon P_{\sigma}) -\pi^i \}.
\end{equation}

As a matter of fact one has even a generalization of 
the main idea of Einstein in gravity. Namely, in GR
we can separate the Universe into two parts---
gravitational field and ``matter'' [3]. In the 
framework of GR the gravitational field is ``dissolved'' 
in the geometry of spacetime. In SuperR all matter is 
``dissolved'' in the geometry of the projective Hilbert space. Therefore we have a consistent approach to
the problem of the divergences, since the spacetime 
localization has, from this point of view, a dynamical character [1,2]. We can exemplify this point in QED. 

The regularization
procedure is effectively the procedure of a 
``delocalization'' of a point-charged electron.  
We do not know, however, the mechanism for the 
suppression of processes of higher orders and it 
is very difficult to find some physically acceptable
mechanism for keeping  the extended
electron from the flying apart [5]. But on closer examination
we will probably find that this difficulty is not
a real one; in the ``geometry of the deformation
of the pure quantum states'' CP(N-1) there are
an absolutely natural stabilizing Goldstone and Higgs mechanism [1,2] (which 
require investigations in detail). It seems much  
better to think of ``deformation'' of the quantum state 
and then looking for localizable solutions (``droplet'')
of some non-linear wave equation 
as a model of non-local quantum particles than 
to begin with the point-charged electron.

\section{Superrelativity and Gravity}
One obtains a nonlinear Klein-Gordon equation (NLKG)
for the effective deformation of quantum state by
requiring that the evolution should move the quantum state along geodesic curve in CP(N-1) [1,2]. 
This equation is as followes:  
\begin{equation}
\Box \Psi^*+ \Box A^* +\Psi^*_{\mu \mu}\frac{\delta A_{\mu}}
{\delta \Psi_{\mu}}+\Psi^*_{\mu}\frac{\delta A_{\mu \mu}}
{\delta \Psi_{\mu}} +\alpha^2(\Psi^*+\Delta\Psi^*+
\Psi^*\frac{\delta\Delta\Psi}{\delta \Psi})=0, c.c.,
\end{equation}
where $A_{\mu}=\frac{\partial \Delta \Psi}
{\partial x^{\mu}}$,$A_{\mu \mu}=\frac{\partial \Delta \Psi}
{\partial x^{\mu} \partial x^{\mu}}$, $\Psi_{\mu}=\frac{\partial \Psi}
{\partial x^{\mu}}$,$\Psi_{\mu \mu}=\frac{\partial \Psi}
{\partial x^{\mu} \partial x^{\mu}}$, and
\begin{equation}
\Delta \Psi^i=- \frac{g\Psi^0\tau^2}
{\sqrt{1+\frac{|\Psi^0|^2}{R^2}}}\Gamma^i_{km}\xi^k \Psi^m. \end{equation}
We seek a solution for the equation in the approximate
form $\Psi=\Phi + \tau \Delta \Phi$ where $\Phi$ obeys
Klein-Gordon equation.
That is, we assume that the solution of the NLKG may be 
represented as a solution of the ordinary
Klein-Gordon equation plus some extra terms arising
from the geometric gauge ``potential'' in CP(N-1)
\begin{equation}
\Gamma^i_{kl} = -2 \frac{\delta^i_k \pi^{l*} + \delta^i_l \pi^{k*}}
{R^2 + \sum_s^{N-1} |\pi^s|^2}. 
\end{equation}
It is clear that we can not hope find a (non-perturbative)
soliton-like solution of the NLKG. But our aim now is
to establish locally a relationship between the spacetime 
structure and the curvature of the projective Hilbert 
space. In order to do it let us look at the equation
for a scalar field in curved spacetime, i.e.,
\begin{equation}
(-g)^{-1/2}\partial_\mu[(-g)^{1/2}g^{\mu \nu} 
\partial_\nu \Phi]+ m^2 \Phi+\eta \varrho(x)\Phi=0,
\end{equation}
where $\eta$ is a coupling parameter, $g^{\mu \nu}$ is 
the metric tensor and $\varrho(x)$ is the scalar 
curvature of spacetime (see for example Ref. [6]). 
One may think of to extra
terms in (2.1) as associeted with a  scalar field 
in a Riemannian geometry as in (2.4). I have obtained
the coefficients in (2.1) in a ``CP(2)-approximation'' 
up to second order in $\tau$ with the help of a program 
in the ``Mathematica''. They have 
a very simple structure but many terms. If one tries
to identify some terms with the Fourier components of
the metric tensor $g^{\mu \nu}$ then one can not be 
certain that different terms in (2.1) are the correct
Fourier components of the scalar curvature $\varrho(x)$
appearing in (2.4).
Notwithstanding, we can think of an ``effective 
Riemannian
geometry'' of the spacetime in which fluctuations 
could be
effectively described by the phenomenological 
parameters $m$ and 
$\eta$. The NLKG equation, as distinct from (2.4), 
contains only one free parameter, the sectional 
curvature $1/R^2$
of the projective Hilbert space. Note  that 
NLKG contains a term with the fine structure constant 
$\alpha$ instead of the mass of the scalar field. This
is a consequence of the choice of the ``classical 
radius'' of meson $r_0=\frac{e^2}{mc^2}$ as unit of 
our scale [1,2]. Such a choice is useful since the inequality
\begin{equation}
\sqrt{\frac{\hbar G}{c^3}}<\frac{e^2}{mc^2}<
\frac{\hbar}{mc}<\frac{\hbar}{\sqrt{2m(E-U)}}
\end{equation}
may be rewritten as follows:
\begin{equation}
\frac{m}{e}\sqrt{\frac{G}{\alpha}}<1<1/ \alpha
<\frac{mc}{\alpha \sqrt{2m(E-U)}}.
\end{equation}
This shows that besides de~Broglie envelope, long-range  
plane waves which depend on modulation by the ``external''
parameters $E,U$, there are a wide range of ``internal'' 
oscillations. These oscillations are connected with 
internal degrees of freedom and should be related to
the spatial destribution of a matter carrier for these
degrees of freedom. The expression for the mass 
distribution can be obtained under the above
mentioned assumption on ``effective Riemannian 
geometry''.
We have, in accordance with the Einstein expression for 
$g_{00}$ [3],
\begin{equation}
g_{00}=1-\frac{2GM}{c^2r}=1-\frac{r_S}{r}=1+\Omega(x,R), 
\end{equation}
where $\Omega(x,R)$ is the collection of terms in 
the decomposition of (2.1) which corresponds to the
time component of the Laplacian. Then one has a formula
for the spatial distribution of mass
\begin{equation}
M(x,R)= - \frac{c^2 r}{2G}\Omega_0(x,R). 
\end{equation}
The implication of this result and some corrections to the nuclear
potential of Yukawa will be discussed elsewhere. 
\section{Discussion}
The unified structure of a ``deformation'' of the pure
quantum state gives us the possibility to investigate  
some general properties of quantum systems. The 
non-trivial
metric and topology of the projective Hilbert space
presumably may endow global solutions of non-linear wave
equation wiyh interesting physical properties.
The tangent fiber bundle of the quantum state space 
in our model is
the main tool of the unified description of matter 
fields in the spirit de Broglie-Schr\"odinger-Bohm.  
The connection in the projective Hilbert space
is a generalization of the well known Panchratnam
connection. This connection defines a parallel 
transport in CP(N-1). A comparison of ``directions'' in
the original spacetime is reduced to the comparison
of field configurations (shapes of ``ellipsoid of
polarization'') by parallel transport projective
Hilbert space. Spacetime structure therefore appears
to arise only ``effectively'' and
problem of localization may be solved in a dynamical
manner (as illustrated  by (2.8)).

\vskip 2cm

ACKNOWLEDGMENTS

I thank Yuval Ne'eman and Larry Horwitz  for numerous 
useful discussions and Yakir Aharonov for his attention 
to this work. This research was supported in part 
by grant PHY-9307708 of the National Science Foundation, 
and by grant of the Ministry of Absorption of Israel.

\end{document}